\documentclass[11pt]{article}

\usepackage{geometry}
\usepackage{currfile}
\usepackage{textcomp}
\usepackage{url}

\usepackage{xcolor,soul}
\sethlcolor{yellow}

\usepackage[pdftex]{graphicx}
\DeclareGraphicsExtensions{.pdf,.jpeg,.png}

\date{1 June, 2020}

\begin{document}

\title{\Large{Wallet Attestations for Virtual Asset Service Providers\\ and Crypto-Assets Insurance\\
~~\\
{\large (Extended Abstract)} }   \\
~~}
\author{
\large{Thomas~Hardjono~~~Alexander~Lipton~~~Alex~Pentland}\\
\large{~~}\\
\large{MIT Connection Science \& Engineering}\\
\large{Massachusetts Institute of Technology}\\
\large{Cambridge, MA 02139, USA}\\
\large{~~}\\
\small{{\tt hardjono@mit.edu}~~{\tt alexlip@mit.edu}~~{\tt pentland@mit.edu}}\\
\large{~~}\\
}

\maketitle

\begin{abstract}
The emerging virtual asset service providers (VASP) industry currently
faces a number of challenges related to the Travel Rule,
notably pertaining to customer personal information,
account number and cryptographic key information.
VASPs will be handling virtual assets of different forms,
where each may be bound to different private-public key pairs on the blockchain.
As such,
VASPs also face the additional problem of the management of its own keys and
the management of customer keys that may reside in a customer wallet.
The use of {\em attestation technologies} as applied to wallet systems
may provide VASPs with suitable evidence relevant to the Travel Rule
regarding cryptographic key
information and their operational state.
Additionally, wallet attestations may provide crypto-asset insurers
with strong evidence regarding the key management aspects
of a wallet device,
thereby providing the insurance industry with
measurable levels of assurance that can become the basis
for insurers to perform risk assessment on crypto-assets bound to keys in wallets,
both enterprise-grade wallets and consumer-grade wallets.

\end{abstract}

\newpage
\clearpage

{\small 
\tableofcontents
}

\newpage
\clearpage


\section{Introduction}

The emerging virtual asset service providers (VASP) industry currently
faces a number of challenges related to the Travel Rule,
notably in connection to customer personal information,
account number and cryptographic key information.
Given that VASPs will be handling virtual assets of different forms,
where each may be bound to a different private-public key pairs on the blockchain,
VASPs also face the additional problem of key management generally.
The key management aspects pertain not only to VASPs themselves
-- as owners of their private-public keys --
but also to customers who posses {\em wallets} (non-custodial)
that hold their private-public keys.
Most end-user consumers have never had to ``handle'' raw keys
or own wallets~\cite{Newman2020}.
From the perspective of the Travel Rule
the private-public keys in customer wallets
become a concern to a VASP when the customer
associates their public-key with their account at the VASP,
but then use the private-key for direct wallet-to-wallet transactions.

In cases where a VASP becomes a custodial of a customer's private-public key,
then the VASP must apply the same degree of protection
as it does for its own keys.
That is, the key management lifecycle~\cite{NIST-800-57} must be a core part
of the cyber-security strategy of the VASP.
The cyber-resilience a VASP's key management infrastructure
has an impact on the business of being a VASP,
including obtaining insurance for
the virtual assets in its possession.
Reports of successful hacks on cryptocurrencies~\cite{Shane2018,Reddy2019,Reiff2019}
has the potential to tarnish the VASP industry as a whole.
Thus, a move towards a new decentralized economy~\cite{Pentland2020a,TradecoinRSOS2018}
brings with it new challenges arising from the need
to decentralized computing infrastructures,
and the need to understand ``trust'' through a decentralized lens.

The recent FATF Recommendations treats virtual assets
as a digital representation of value,
and as such is covered under the existing Anti-Money Laundering (AML)
regulations and the Travel Rule.
This means VASPs must be able to obtain, verify, retain and share
respective customer information (originator and beneficiary)
in the same manner as existing financial institutions.
This means, among others, personal information,
account information, and transactions information.
However, as pointed out in~\cite{HardjonoLipton2020a},
in the case of virtual assets on a blockchain,
the key-ownership information and key-operator information
becomes another aspect of the customer's account that
VASPs must manage.
This is because virtual assets on blockchains are directly controlled
by cryptographic keys, 
and therefore by the entity who controls the private-public key pair.

In this paper, we seek to address technological means
to provide the ownership information for private-public keys that are located 
within a customers wallet device.
More specifically, 
we explore how {\em attestation technologies} as applied to wallet systems
can provide VASPs with means to prove operational controls of keys in wallets.
The capability to obtain ``visibility'' into the state of keying material in
a wallet device -- without revealing the customer's private-keys --
provides a good basis for a VASP to perform management (remote management)
of the customer wallet.
This in turn allows VASPs to address some of the regulatory compliance 
requirements arising from the Travel Rule.

This paper is arranged in the following manner.
We discuss the Travel Rule in
Section~\ref{subsec:TravelRule},
and discuss some of the corresponding
challenges to VASPs in Section~\ref{sec:Challenges-VASP-Industry}.

In Section~\ref{sec:TCGAttestation} we review the concept of device attestations
in the context of the VASPs and blockchains use-case,
as a means to assist a VASP in obtaining better visibility into
the internal state of the wallet devices~\cite{TCG-Glossary-2017,NIST-800-193}.
We review the current efforts in industry relating to the standardization of attestation architectures
and evidence conveyance protocols.
We discuss the application of device attestations to wallets in Section~\ref{sec:AttestWallets},
and explore how communities of VASPs arranged in a consortium or ``trust network''
can use of shared attestation services
as a means for VASPs to collectively address some of the challenges
arising due to the Travel Rule.

In the current work,
we seek to make the concepts around device attestations to be more easily
understandable and more accessible
to the broad readership interested in VASPs, wallets and the virtual assets industry generally.
Much of the language describing attestations come from
the area of trusted computing,
which has been heavily influenced by the two-decades of the 
development of the Trusted Platform Module (TPM) chip~\cite{TPM1.2specification,TPM2.0specification}.
As such, we will strive to abstract-up from the various design 
features of the TPM and focus on the intent of some of these features, 
narrowing our interest on those features that support attestation 
and its potential use for VASPs in their ecosystem.
Readers interested in details of the TPM are directed 
to the excellent works of~\cite{Proudler2002,ChallenerYoder2008,Proudler2014}.

%
%
%
%

\section{Virtual Assets, VASPs and the Travel Rule}
\label{subsec:TravelRule}

Since the advent of the Bitcoin system in 2008~\cite{Bitcoin},
there has been an ever-growing interest among the general public
in the potential use of crypto-currencies (``crypto'') and virtual assets
for decentralized unmediated financial transaction.
One of the key issues in this nascent crypto industry
is the need for  virtual asset service providers
to comply with the various financial regulations
related to Anti-Money Laundering (AML), terrorism financing
and other banking related regulations.
At the international level,
the inter-governmental body established to 
set the standards and promote the effective
implementation of legal, regulatory and operational measures is
the Financial Action Task Force (FATF)~\cite{FATF-website}.

\subsection{FATF Recommendations No.~15 and the Travel Rule}

A major milestone event in the crypto-currencies industry was
the publication of the FATF Recommendation {No.~15} in late 2018
which provided a comprehensive definition the notion of {\em virtual Assets} and 
{\em virtual Asset Service Providers} (VASP)~\cite{FATF-Recommendation15-2018}:
\begin{itemize}

\item	{\em Virtual Asset}: A virtual asset is 
a digital representation of value that can be 
digitally traded, or transferred, and can be used for payment or investment purposes. 
Virtual assets do not include digital representations of fiat currencies, 
securities and other financial assets that are 
already covered elsewhere in the FATF Recommendations.

\item	{\em Virtual Asset Service Providers} (VASP): Virtual asset service provider means 
any natural or legal person who is not covered elsewhere under the Recommendations, 
and as a business conducts one or more of the following activities or 
operations for or on behalf of another natural or legal person:
(i) exchange between virtual assets and fiat currencies; 
(ii) exchange between one or more forms of virtual assets;
(iii) transfer of virtual assets;
(iv) safekeeping and/or administration of virtual assets or instruments enabling control over virtual assets; and
(v) participation in and provision of financial services related to an issuer's offer and/or sale of a virtual asset.
\end{itemize}
In this context of virtual assets, transfer means to conduct a transaction 
on behalf of another natural or legal person that moves 
a virtual asset from one virtual asset address or account to another~\cite{FATF-Recommendation15-2018}.

One of the main requirements called out in the FATF Recommendation {No.~15}
and its accompanying Guidelines document~\cite{FATF-Guidance-2019}
is the mandated need for
VASPs to retain information regarding the originator and beneficiaries of
virtual asset transfers.
That is, a VASP must posses accurate information regarding
a customer before aiding that customer in conducting transactions in virtual assets.

Another important implication of the Recommendation {No.~15} is that cryptocurrency
exchanges and related VASPs must be able to share the originator and
beneficiary information for virtual asset transactions. 
This process -- also known as the funds {\em Travel Rule} -- originates from the US Bank Secrecy Act
(BSA~31 USC Secs. 5311-5330), which mandates that financial institutions
deliver certain types of information to the next financial institution
when a funds transmittal event involves more than one financial institution.
This rule became effective in May 1996 and was issued by the U.S. Treasury
Department's Financial Crimes Enforcement Network (FinCEN).
Other groups of financial institutions (e.g. Wolfsberg Group~\cite{Wolfsberg2012})
have also tailored their AML principles based on the FATF Recommendations.

\subsection{Key Management Configurations}
\label{subsec:KeyManagementConfigurations}

In contrast to traditional banking institutions
-- which have been operating under the same Travel Rule for over two decades --
VASPs today have the additional problem of dealing with cryptographic keys
associated with customers.
This stems from the fact that virtual assets on a blockchain
is directly controlled through the private-public key
associated with that virtual asset.

\begin{figure}[t]
\centering
\includegraphics[width=1.0\textwidth, trim={0.0cm 0.0cm 0.0cm 0.0cm}, clip]{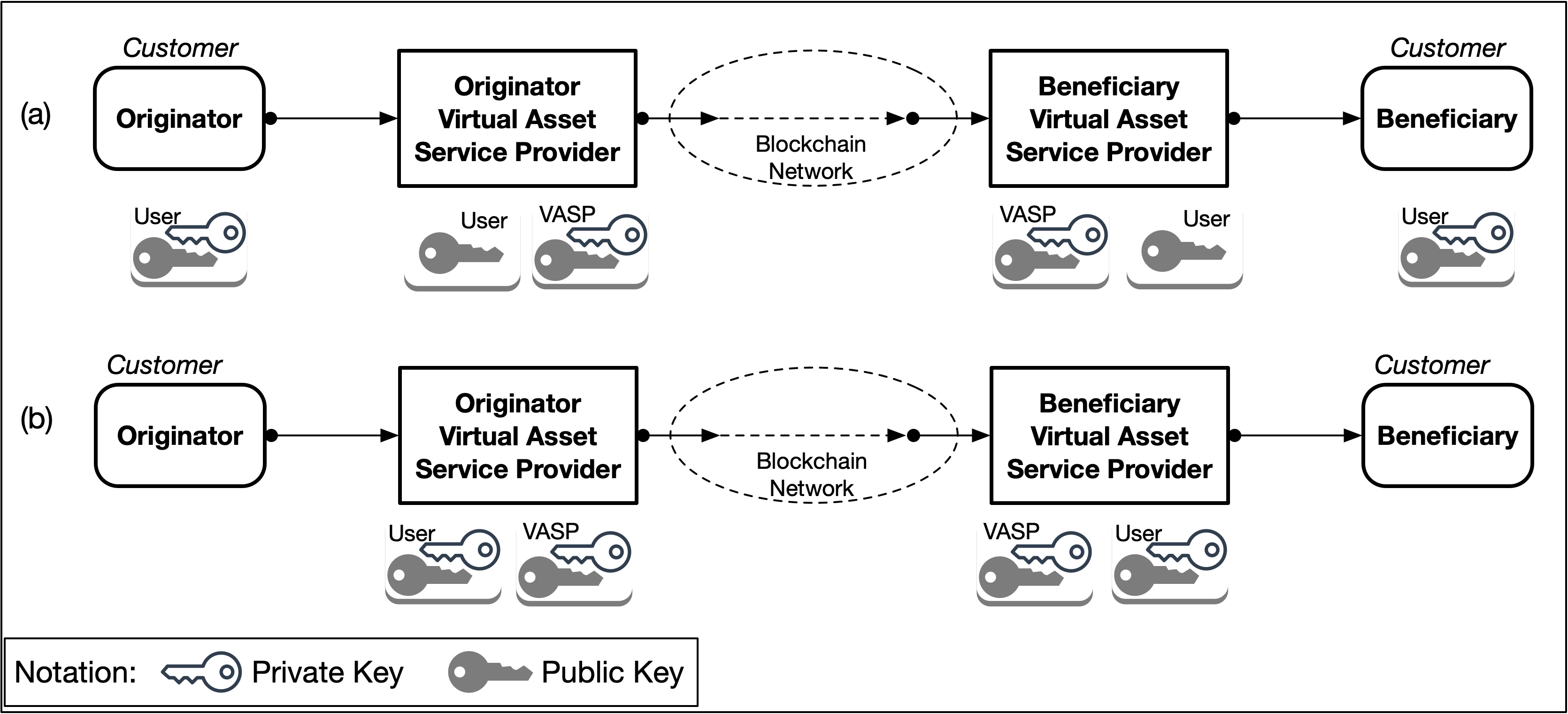}
\caption{Two possible VASP cryptographic key management configurations (after~\cite{HardjonoLipton2020a})  }
\label{fig:walletconfigs}
\end{figure}

For VASPs there are a number of possible models or configurations with regards
the management and use of the private-public key pair used to sign transaction 
on the blockchain.
Two of the configurations relevant to the current discussion
are shown in Figure~\ref{fig:walletconfigs}
(see~\cite{HardjonoLipton2020a} for other variations).

In configuration (a) of Figure~\ref{fig:walletconfigs},
the customer holds its private-public keys in the customer's wallet.
The Originator-VASP holds a copy of the customer's public key,
possibly enveloped within a digital certificate 
(e.g. {X.509} certificate~\cite{rfc2459,rfc5280,ISO9594-pubkey}).
There are at least two ways involving VASPs
for a wallet-based key-pair to be used
by the originator customer to transfer virtual assets to a beneficiary.
In the first case,
the originator creates the signed transaction (addressed to the beneficiary)
and delivers to its VASP (Originator-VASP),
requesting the VASP to transmit the virtual asset onto the blockchain.
This provides the opportunity for the VASP to perform the relevant
Travel Rule verifications with regards to the destination beneficiary.
Thus, technically speaking the VASP acts similar to a forwarding ``gateway''
that processes ``ready-to-transmit'' transactions signed by the private-key in the customer wallet.
In the second case,
the originator customer is the entity actually transmitting the transaction 
(i.e. direct from its wallet).
However, before doing so the the originator relies on its VASP
to perform the Travel Rule verifications regarding the destination beneficiary.
Then, once the originator customer obtains permission (``green light'') from its VASP,
the originator transmits the transaction directly from its wallet.
In either of these cases, the Travel Rule applies to the VAPSs,
the originator and the beneficiary.

In configuration (b) of Figure~\ref{fig:walletconfigs}
the customer does not hold any private-public keys or own any wallet.
Instead, the customer opens an account at the VASP  
and all the customer's asset-transactions are dealt with by the VASP.
In this approach,
the VASP has at least two options with regards key management.
In the first option, the VASP could hold a separate private-public key pair
for each customer and employ that key-pair on behalf of the customer when
transacting the customer's virtual assets.
This approach is often referred to as the {\em key custodial} configuration~\cite{GDF2019-Custodial}.
In the second option, the VASP employs its own private-public key pair
to sign all asset-related transactions.
In this case various customer assets can be said to be {\em commingled}.
With commingled assets/accounts,
the VASP can batch together multiple transactions from its various customers,
thereby reducing the overall cost (fees) of the transaction.

\subsection{Customer Accounts \& Key Ownership Information}
\label{subsec:KeyOwnershipInformation}

Aside from the customer personal or business information,
the Travel Rule also requires that financial institutions and VASPs
furnish the account numbers when performing asset transfers.
This brings a number of interesting challenges
with regards to how ``account numbers'' are to be realized
in the case of virtual assets on the blockchain and how private-public keys
are to be associated with accounts.

With regards to account numbers,
the work of OpenVASP~\cite{OpenVASP-2019} has proposed the establishment of a standardized
{\em VASP code} for each VASP consisting of the last 32-bits of the public-key of the VASP.
Although the proposal is expressed in the context of the Ethereum system~\cite{Buterin2014},
the notion of a 32-bit VASP code is generally speaking logical, practical and indeed necessary
from a VASP operations point of view.
The 32-bit VASP code provides sufficient bit-space to uniquely identify up to millions of VASPs in the future.
When used in the context of Ethereum,
VASPs also have the option of claiming a namespace within the Ethereum Name Service (ENS),
and thus allowing the VASPs public-keys 
(i.e. its customer's public-keys) to be known and discoverable within the ENS context.
The proposal of~\cite{OpenVASP-2019} also includes the notion
of a unique {\em Virtual Asset Account Number} (VAAN)
for each customer,
where the first 32-bit of the VAAN number is the VASP code.
This proposal is logical and reasonable,
and it mimics
the familiar bank ABA routing numbers and account numbers.

Given that private-public key pairs are the means to control virtual assets,
we believe that for compliance to the Travel Rule,
VASPs will also need to maintain information regarding the key-pairs of their customers.
VASPs most likely will be required
to share key-related information or certificates
with other VASPs and financial institutions involved in virtual asset transfers.
Key-related information should be considered as another
attribute of the customer account information that is required in order
to comply to the Travel Rule.

Thus, in the context of the various possible key management configurations 
(Section~\ref{subsec:KeyManagementConfigurations} and Figure~\ref{fig:walletconfigs}),
it is useful to distinguish between {\em key-ownership information} 
and {\em key-operator information}~\cite{HardjonoLipton2020a}.
This is particularly important for VASPs
from a risk management and assets-insurance perspective~\cite{John2018,KharifLouis2018}.
\begin{itemize}

\item	{\em Key ownership information}: This is information pertaining to the 
legal ownership of cryptographic public-private keys.
The traditional mechanism to denote legal ownership
is through the registration of the public-key to
a Certification Authority (CA),
and for the CA to issue a public-key certificate for that key~\cite{eSign2000}.
The certificate itself is signed by the CA,
effectively making the CA a notary asserting the ownership of the key-pair.
The CA itself must be a registered business
operating under a service contract
(known as the Certificate Practices Statement).

It is worth noting that ability for an entity to prove possession of a private-key
-- such as exercising the private-key in a challenge-response protocol 
(e.g. CHAP~\cite{rfc1994}) or
in signing a transaction on the blockchain --
does not imply legal ownership of the key.

\item	{\em Key operator information}: 
This is information or evidence pertaining to the legal custody by a VASP
of a customer's public-private keys.
This information is relevant for VASPs which
adopt a key-custodial business model, 
where the VASP holds and operates the customer's public-private keys
to sign transactions on behalf of the customer.
Here, the customer legally owns the public-private key-pair,
but the customer-authorized legal operator of the key-pair is the VASP.

\end{itemize}

In the next section we review a number of challenges
faced by the emerging VASP industry.
Some of these challenges are related to the Travel Rule,
while others are related to the operational aspects of VASPs
and the need for VASP industry to develop and establish
new trust infrastructures to support transaction
on decentralized blockchain networks.

\section{Current Challenges in the VASP Industry}
\label{sec:Challenges-VASP-Industry}

Aside from crypto-asset insurance issues,
there are currently a number of challenges faced by the VASP community globally.
These challenges arise not only because of existing regulations
with regards to AML/FT, but also because
the of the rapid changes occurring in blockchain and DLT technologies.
We briefly discuss some of these issues
as a background to the discussion on attestations in the ensuing sections,
and later to a discussion on the benefits of wallet attestations 
to VASPs and to crypto-asset insurers.

In the following we use the term {\em regulated wallet}  
to denote a wallet system (hardware and software) that is in possession
of a customer of a supervised (regulated) VASP~\cite{FINMA-Guidance-2019}.
The understanding is that the customer's private-public keys
are within the wallet device, and that the wallet device is
in the possession and under the control of the customer
(Figure~\ref{fig:walletconfigs}(a)).
From the perspective of customer information
we use the term ``regulated'' in the sense of FINMA~\cite{FINMA-Guidance-2019}.
This means that the customer is registered at a VASP,
owns an account at the VASP, 
and the VASP is able to fulfill the compliance requirements of the Travel Rule
for that customer.

We use the term {\em private wallet} to denote
a wallet system belonging to an {\em unverified entity}.
This implies that in the extreme case
the wallet-holder information is unattainable by a VASP,
despite the VASP querying other VASPs in the messaging network
and querying certification authorities (CA) reachable by the VASP.

\subsection{A Secure Messaging Network for VASPs}
\label{sec:MessagingNetwork}

The need to exchange customer account information and other private information
raises the question regarding the need for a secure {\em messaging network} 
for VASPs~\cite{Hardjono2019b,TRISA-2019}.
Such a messaging network may or may not be implemented using blockchain technology.
However, there are a number of 
fundamental requirements for such a messaging network.
Among others, it must:
(i) protect customer/VASP data privacy,
(ii) provide secure transport of customer/VASP information between endpoints,
(iii) operate based on the strong identification and authentication of VASPs in the network,
and
(iv) be able to operate independent of any specific virtual-asset blockchain network,
current or future.

The need for a secure transport (e.g. SSL/TLS connection) from one VASP to another
implies that VASPs will also need to posses private-public keys
designated for negotiating the secure transport.
It is good security practice to keep
the SSL/TLS private-public keys distinct
from the private-public keys
used by a VASP to sign transactions for the blockchain. 
Ideally, the SSL/TLS private-public keys should be wrapped
in a digital certificate (e.g. X.509),
possibly using Extended Validation (EV) certificates that carry business information
regarding the VASP that owns the EV-Certificate~\cite{CAB-Forum2020}.

\begin{figure}[t]
\centering
\includegraphics[width=1.0\textwidth, trim={0.0cm 0.0cm 0.0cm 0.0cm}, clip]{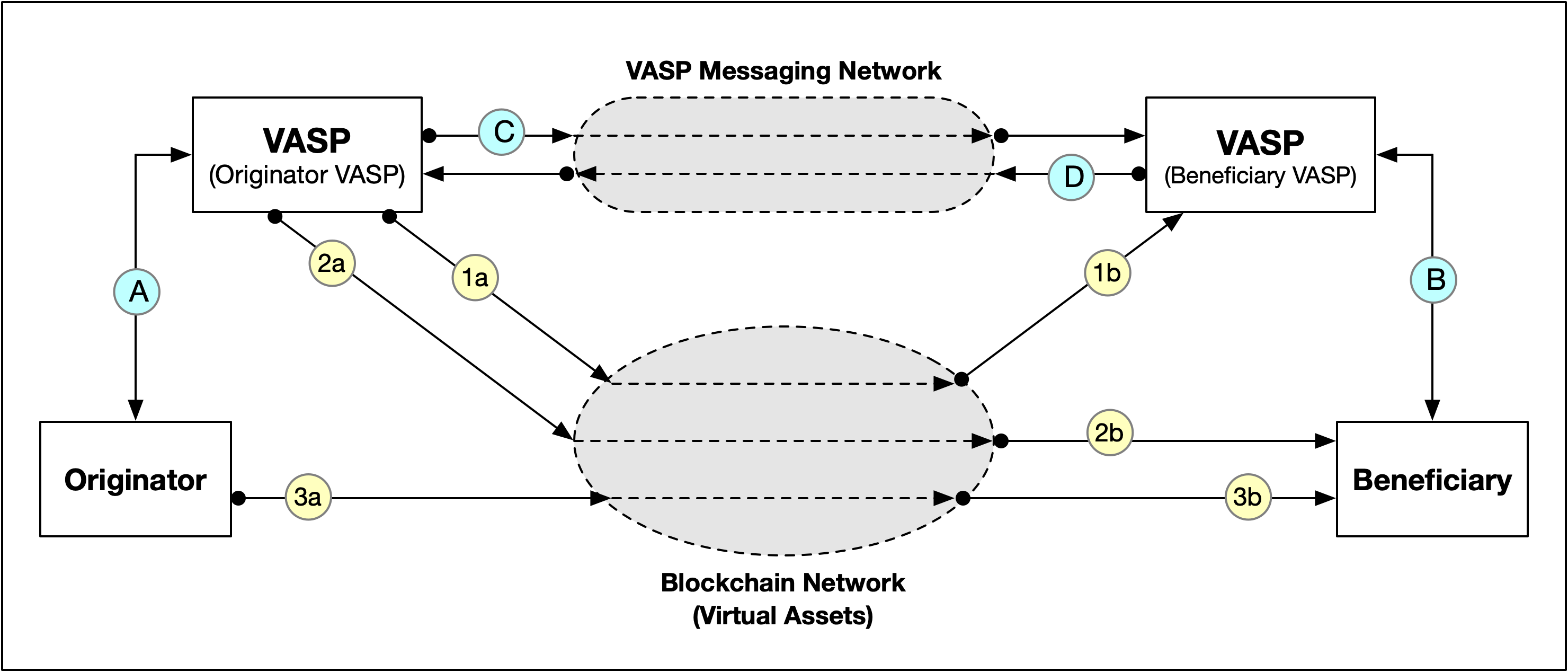}
\caption{The VASP Messaging Network and Blockchain Network}
\label{fig:VASPMessagingNetwork}
\end{figure}

\begin{figure}[t]
\centering
\includegraphics[width=0.8\textwidth, trim={0.0cm 0.0cm 0.0cm 0.0cm}, clip]{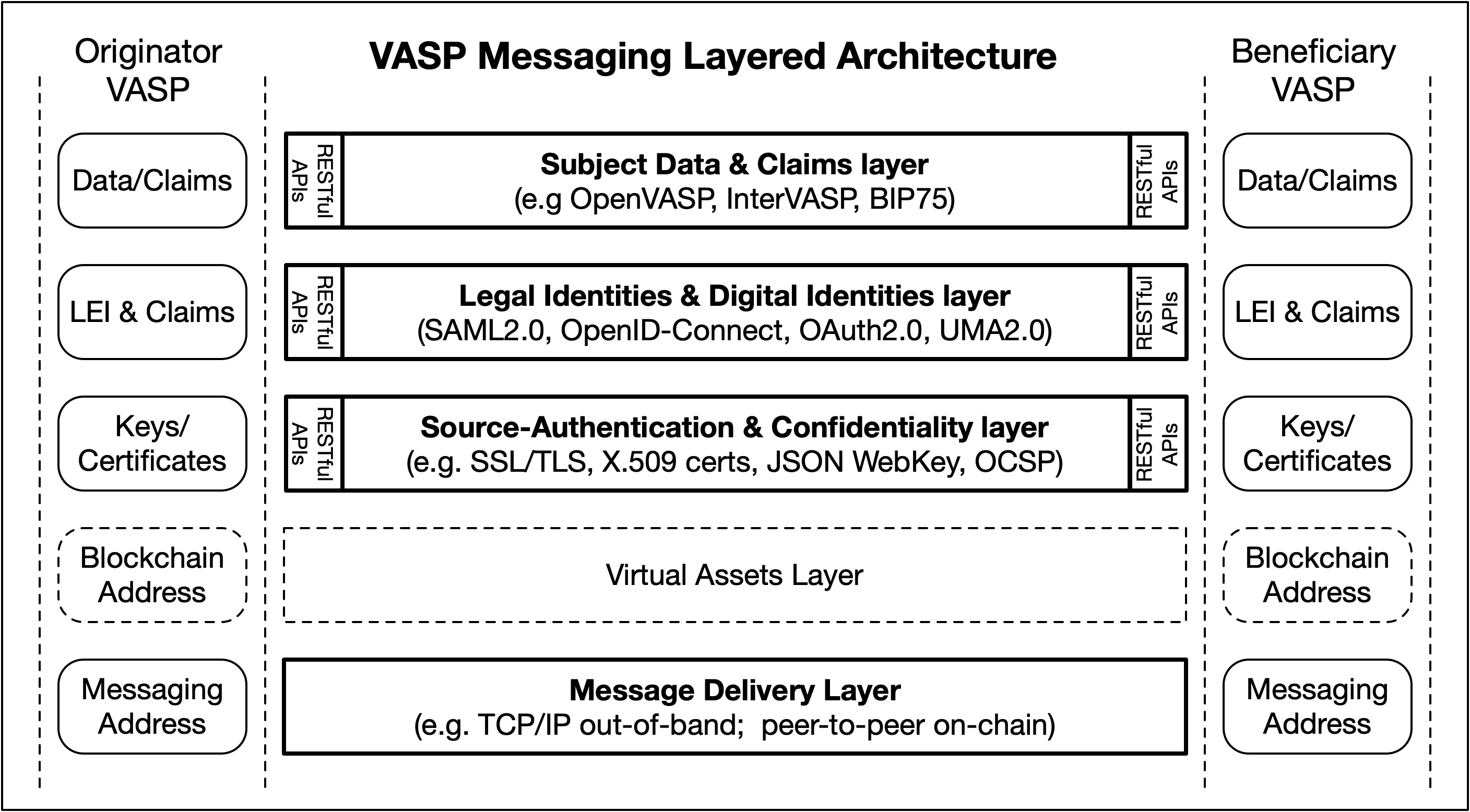}
\caption{A layered architecture for the VASP Messaging Network}
\label{fig:LayeredMessagingNetwork}
\end{figure}

Such a messaging network must be able to evolve and persist over the next few decades,
independent of (but informed by) the technological advancements in blockchains.
It must allow a community of VASPs to exchange information pertinent to the Travel Rule,
for any type of virtual asset transfers 
(e.g. crypto-currencies, tokenized assets, etc)
on current and future blockchain systems.
Figure~\ref{fig:VASPMessagingNetwork} provides a high level illustration
of a VASP messaging network, shown to be logically separate from
the virtual asset blockchain.
Figure~\ref{fig:LayeredMessagingNetwork} illustrates
a possible layered architecture for the VASP messaging network.

In Figure~\ref{fig:VASPMessagingNetwork}
the Originator (Beneficiary) customer is assumed 
to have a business relationship (e.g. account) with
the Originator-VASP (Beneficiary-VASP).
This is shown as (A) and (B).
The VASP messaging network is denoted as (C) and (D),
where the Originator-VASP can securely transmit the originator customer
information to the Beneficiary-VASP in (C),
and vice versa in (D).
 
Efforts are currently underway to develop such a messaging network
and the related trust infrastructures (e.g. certificates, decentralized directories)
to support VASPs in complying to the various aspects of 
the Travel Rule (see~\cite{TRISA-2019,OpenVASP-2019}).
A standard customer information model has recently been developed~\cite{InterVASP2020FINAL} 
that would allow VASPs to interoperate with each other with semantic consistency
when exchanging customer data.

\subsection{Synchronization of Customer Information with Transactions}
\label{subsec:Synchronization}

The Travel Rule requires VASPs to ``synchronize'' (track or correlate)
between transactions on the virtual assets blockchain with customer information.
More specifically,
this means that both the Originator-VASP and the Beneficiary-VASP
must account for every customer transaction,
and correlate transaction-related information
to the correct originator and beneficiary respectively.

Related to this is the question of customer data privacy~\cite{GDPR}.
Every instance of the exchange of customer information
must be driven by (pertain to) transactions from the customers.
The corollary is that
VASPs are not permitted to exchange customer-related information outside 
the context of a customer transaction.

This introduces a number of complications for VASPs
with regards to the latency in the completion of customer transactions.
A given VASP may need to delay the transmission of customer transaction until
the relevant Travel Rule information (ie. regarding the beneficiary)
has been obtained and verified.
This requirement is bi-directional or mutual between VASPs.
This means that the Beneficiary-VASP must also obtain and verify 
the originator customer information (e.g. from the Originator-VASP)
before accepting transfers of virtual assets from the originator or its Originator-VASP.
Such delays may be deemed to be ``disappointing'' by 
customers accustomed to media crypto-hype~\cite{FurlongerKandaswamy2018,LitanLeow2019}.
Solving this challenge is beneficial for all VASPs around the world,
but will require corporation among competing VASPs.

\subsection{Direct Transactions from Wallets}
\label{subsec:TransactionRegWallet}

One of the dilemmas faced by VASPs is the desired (demand) on the part of customers
to perform direct transactions from the customer's wallet (i.e. peer-to-peer).
This is illustrated in Figure~\ref{fig:VASPMessagingNetwork} in flow~(3a) and flow~(3b).
This demand can be addressed if both wallets (originator and beneficiary)
are regulated wallets respectively.
However, the dilemma arises when only the originator wallet is regulated,
and the beneficiary information is not verifiable immediately by the Originator-VASP.

One possible solution is for VASPs to permit (pre-authorize) customers
with regulated-wallets to transact up to a maximum daily limit as defined by local regulations
(e.g. \$3K per day).
The Originator-VASP must then perform the verification of the beneficiary information
{\em after} the transaction has been confirmed on the blockchain.
The task of post-event verification would be made easier if the beneficiary
is found to be associated with a Beneficiary-VASP.

This limited pre-authorized solution could be tightly integrated into
user authentication/authorization {\em Single Sign-On} (SSO) 
protocol~\cite{SAMLcore,RFC6749-Formatted,UMACORE1.0}
via the customer's wallet.
When a customer seeks to perform a direct transaction from its wallet,
the customer/user would need to perform SSO (using the user's wallet) 
to the authorization service of the VASP,
which should just take a few seconds.
This process essentially provides a mechanism for the wallet to
notify the VASP authorization server
that the wallet will soon be transmitting a transaction
in a direct P2P fashion to a beneficiary wallet.
The SSO event provides a window of time for the VASP to verify whether
the beneficiary is a known entity to the VASP
(e.g. previously transacted from the customer)
and whether a Beneficiary-VASP can be located corresponding
to the beneficiary wallet holder.

More sophisticated solutions may be devised
based on well-known database transaction processing principles,
such as the 2-Phase Commit (2PC) protocol~\cite{TraigerGray1979,Gray1981}.
However, this topic is out of scope for the current work.

\subsection{VASP Transactions to Private Wallets}
\label{subsec:PrivateWallets}

Another acute problem pertains to cases
where an originator customer of a regulated VASP
requests asset transfers to an address (public-key) 
of a private wallet.
This is shown as flow~(2a) and flow~(2b) in Figure~\ref{fig:VASPMessagingNetwork}.

The originator customer may only have informal and incomplete information 
regarding the beneficiary holder of the private wallet (i.e. either a person or an organization).
The challenge, therefore, becomes one in which the Originator-VASP needs to seek
information at other VASPs regarding that destination address.
Indeed, this is one of the main reasons VASPs need to create
a trust network or industry consortium operating under a legal trust framework
(see~\cite{HardjonoLipton2020a}).
Efforts such as TRISA~\cite{TRISA-2019} are aimed at solving this dilemma,
based on discovery protocols as well as VASP-level federated directories.

The problem also has data privacy dimension~\cite{GDPR}.
A remote VASP located in a different jurisdiction
may have verified information regarding holder of a private wallet or address.
However, that remote VASP may be prohibited (e.g. under local data privacy regulations)
from disclosing knowledge of
the owner of a given address or public-key.
As such, the remote VASP maybe prohibited from even responding to the query.

\subsection{On-Boarding and Off-Boarding Customers}
\label{subsec:OffBoarding}

There are a number of challenges related to the on-boarding of a customer
possessing a wallet.
In the case that the customer wallet is regulated 
(i.e. previously known to another regulated VASP),
then there are a number of practical issues
that the {\em acquiring} VASP needs to face.
These include:
(i) validating whether prior to on-boarding the wallet was regulated or private;
(ii) validating that the keys present within the wallet corresponds to the customer's
historical transactions (confirmed on the blockchain);
(iii) verifying whether a backup/migration of the wallet has occurred in the past;
(iv) determining whether the customer's assets should be moved to new keys, and if so,
how the ``old'' keys will be archived;
and so on.

In the case where the customer's wallet is private,
then the VASP may choose to require the customer
to create a separate instance of the wallet application
within the device, associated with the VASP.
This provides a way for the VASP to be responsible
only for the ``regulated partition'' of the wallet
while allowing the customer to retain the private portions of the wallet.
Trusted Execution Environment (TEE) technologies
such as Intel SGX~\cite{McKeen2013,McKeen2016}
may provide a way to achieve this partition, and at the same time
provide evidence regarding the partition on the wallet device which is under
the Travel Rule responsibility of the VASP.

The case of a customer leaving a VASP also introduces a number
of questions that may be relevant under the Travel Rule.
The {\em releasing} VASP may need to address various
question, including:
(i) preparing evidence that the wallet was in a regulated state whilst
the owner of the wallet was a customer of the VASP;
(ii) whether the customer's assets should be moved to a temporary set of keys,
denoting the end of the VASP's responsibilities for the customer under the Travel Rule;
(iii) obtaining evidence from the wallet that the ``old'' keys (non-migrateable keys)
have been erased from the wallet device,
thereby rendering the keys unusable in the future by the customer;
and so on.

\subsection{Cross-Jurisdiction Asset Transfers}
\label{subsec:Cross-jurisdiction}

Within certain jurisdictions
the operational requirements may be more stringent.
For example,
in Switzerland the FINMA ordinance on Anti-Money Laundering (AMLO)
makes it unambiguously clear that no exception is permitted
for payments (i.e. virtual asset transfers)
involving ``unregulated wallet providers''~\cite{AMLO-FINMA2015}. 
This rule, among others,
is to prevent (reduce) problems faced by supervised providers (i.e. VASPs)
in cases where it has to deal with 
virtual asset transfers from an unregulated wallet provider.

This brings into focus the challenge
of how virtual asset transfers will occur
between two VASPs where each are operating under different
jurisdictions with different levels of stringencies.
Furthermore,
as long as an institution (i.e. VASP) supervised by FINMA is not able to send and
receive the customer information required in payment transactions, such transactions
are only permitted between wallets of the institution's own customers.

Relevant to the current work on attestations
is the requirement in FINMA that the ownership of a wallet be
proven using ``suitable technical means''~\cite{FINMA-Guidance-2019}.

\section{A Standard Architecture for Attestations}
\label{sec:TCGAttestation}

Recently, the notion of attestations has have garnered interest within
different technical standards organizations and industry consortiums,
beyond the TCG alliance (e.g. FIDO Alliance~\cite{FIDO-key-Attestation-2015}, 
Global-Platform~\cite{GlobalPlatform2012}, IETF~\cite{IETF-RATSWG}).
Despite the notion of device attestations nearing two-decades 
in age~\cite{TPM2003Design},
the concepts around attestations -- such as endorsements, validations and freshness --
are just recently coming into wider attention in the broader industry.
The hope is that a canonical attestation architecture 
will allow standards to be developed that implement
the various protocols and flows  
for relevant sectors and products
(routers and network equipment~\cite{TCG-RIV-2020,IETF-rats-network-device-attestation-05},
mobile devices~\cite{GlobalPlatform2012}, 
cloud stacks~\cite{OpenCompute-website2020}, etc).
By having a common reference architecture,
different efforts can share common terminologies, concepts and implementations
and therefore affect a reduction in costs of developing and deploying
the infrastructures supporting cyber-resilience and trustworthy computing generally.

\subsection{Attestations}

The fundamental idea of attestations of a ``thing'' (e.g. a computing device)
is that of the conveyance of truthful information regarding 
the (internal) state of the thing being attested to.
In the related literature on trustworthy computing 
the term ``measurement'' is used to mean the 
act of collecting (introspecting) claims or assertions 
about the internal state, 
and delivering these claims as evidence to an 
external party or entity for automated review and security assessment.

\begin{figure}[t]
\centering
\includegraphics[width=1.0\textwidth, trim={0.0cm 0.0cm 0.0cm 0.0cm}, clip]{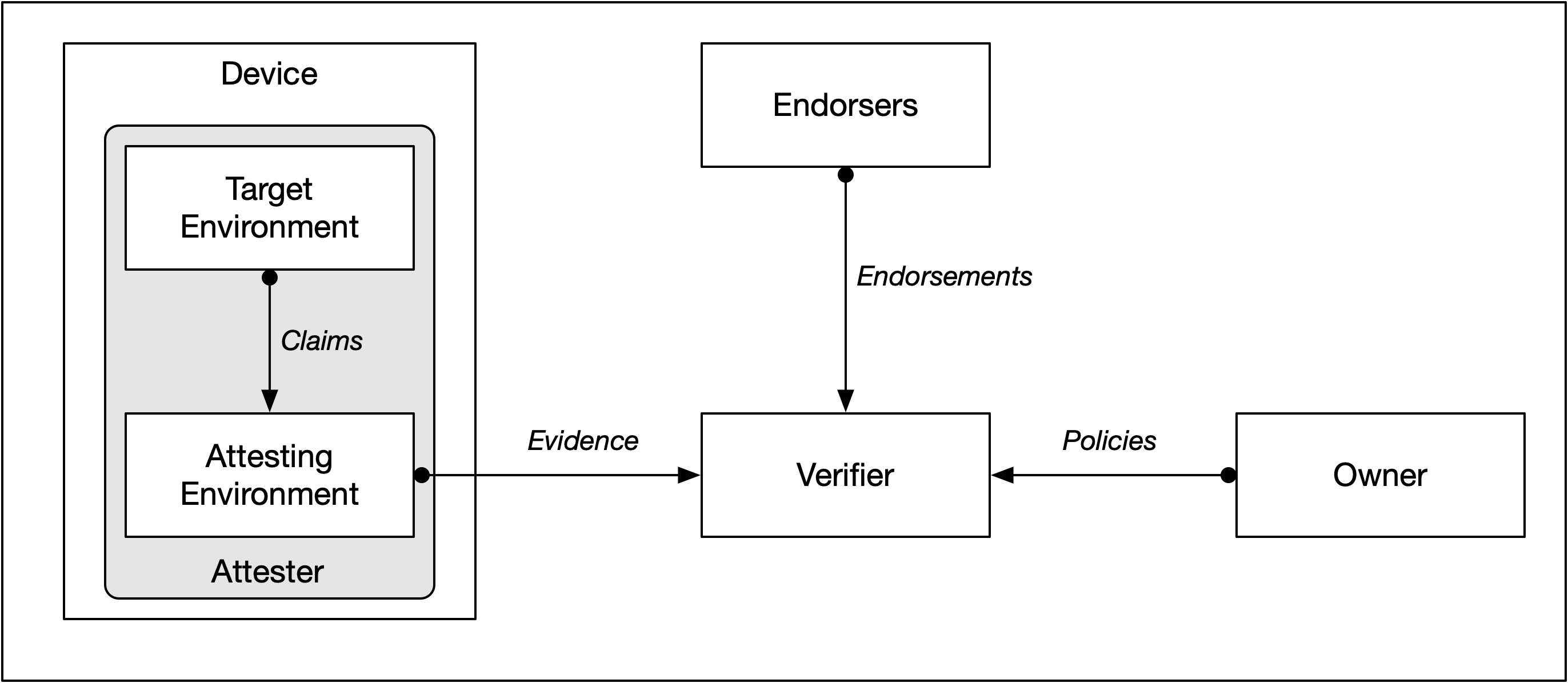}
\caption{Overview of the concept of the Attester and Attesting Environment}
\label{fig:AttestingEnvironment}
\end{figure}

However, as we know today
computing environments can be structurally complex
and may consist of multiple elements (e.g. memory, CPU, storage, networking, firmware, software),
and computational elements can be linked and composed to form computational pipelines, arrays and networks.  
Thus, the dilemma is that not every computational element can be expected to be capable of attestation.
Furthermore, attestation-capable elements may not be capable of attesting every 
computing element with which it interacts.  
The attestation capability could in fact be a computing environment itself (Figure~\ref{fig:AttestingEnvironment}).
The act of monitoring trustworthiness attributes, 
collecting them into an interoperable format, 
integrity protecting, authenticating and conveying 
them requires a computing environment -- one that 
should be separate from the one being attested. 
Figure~\ref{fig:AttestingEnvironment} illustrates the recognition of this distinction,
namely of the {\em target environment} being attested to,
and the {\em attesting environment} that performs 
the work stated above\footnote{An example of an attesting environment is
the Quoting Enclave within Intel SGX~\cite{McKeen2013,McKeen2016}.}.

The complexity of the problem has led to a number of efforts in industry
to define an {\em attestation architecture} that incorporates
some of these key concepts -- such as the concept of the root-of-trust --
and to develop standards that implement attestation concepts~\cite{TCG-ATTEST2020,IETF-RATSWG}.
The roles and functions of the attestation architecture is shown in Figure~\ref{fig:AttestingEnvironment}.
In a nutshell,
in Figure~\ref{fig:AttestingEnvironment} an attester
conveys evidence of trustworthiness (of the attested target environment)
to a verifier entity.
The verifier operates based on policies that are supplied by the owner of the verifier.

\begin{figure}[t]
\centering
\includegraphics[width=1.0\textwidth, trim={0.0cm 0.0cm 0.0cm 0.0cm}, clip]{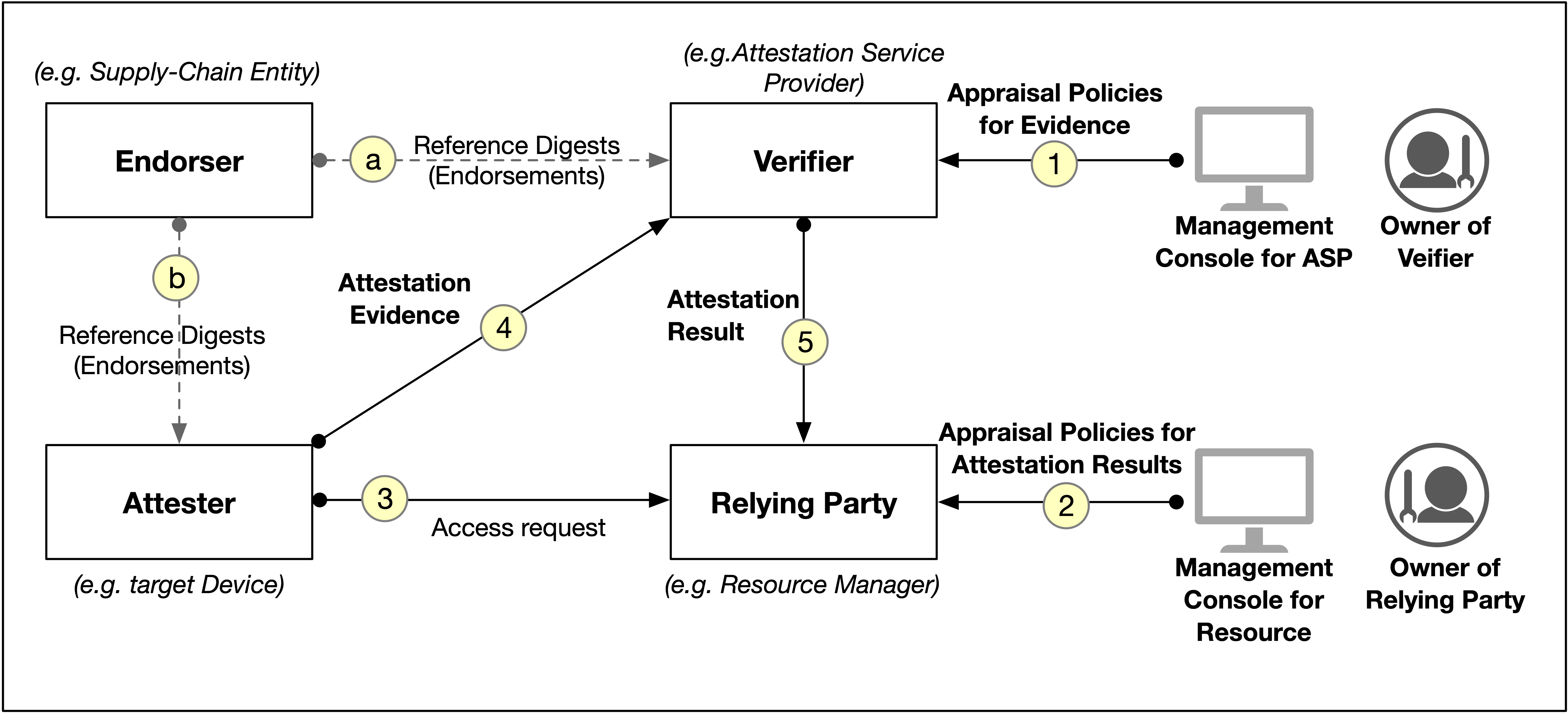}
\caption{Canonical architecture for attestations (after~\cite{TCG-Attestations-Arch2020,rats-arch-02,CokerGuttman2011})}
\label{fig:TCG-Canonical-Model}
\end{figure}

\subsection{Entities, Roles and Actors}
\label{subsec:TCG-Attestation-entities}

The attestation architecture of~\cite{TCG-Attestations-Arch2020}
defines of a set of {\em roles} that 
implement attestation flows.  Roles are hosted by {\em actors},
where actors are deployment entities.  
Different deployment models may coalesce or separate various actor 
components and may call for differing attestation conveyance mechanisms.  
However, different deployment models do not fundamentally modify attestation roles, 
the responsibilities of each role, nor the information that flows between them.
In the following sections, we may use the actor and role terminology interchangeably 
when appropriate in order to simplify discussion (see Figure~\ref{fig:TCG-Canonical-Model}).
\begin{itemize}

\item	{\em Attester}: 
The Attester (e.g. target device) provides attestation Evidence to a Verifier. 
The Attester must have an attestation identity that is used to
authenticate the conveyed Evidence and establishes an attestation endpoint context. 
The attestation identity is often established as
part of a manufacturing process that embeds 
identity credentials in the entity that implements an Attester.

\item	{\em Verifier}:
The Verifier accepts Endorsements (from Endorsers) and Evidence (from the Attester) 
then conveys Attestation Results to one or more Relying Parties. 
The Verifier must evaluate the received Endorsements and Evidences
against the internal {\em appraisal policies} chosen or configured by the owner
of the Verifier~\cite{CokerGuttman2011}.
The Attestation Service Provider (ASP) 
is typically the actor which implements the Verifier role.
An example of the ASP role is described in~\cite{zichardjono2013}.

\item	{\em Relying Party}:
The Relying Party (RP) role is implemented by a resource manager 
that accepts Attestation Results from a Verifier. 
The Relying Party trusts the Verifier to correctly evaluate 
attestation Evidence and Policies, and to produce a correct
{\em Attestation Result}. 
Thus, we assume that the RP and the Verifier has a business relationship
or some other basis for trusting one another. 
The Relying Party may further evaluate Attestation Results 
according to Policies it may receive from an Owner.
The Relying Party may take actions based on the evaluation of the Attestation Results.

\item	{\em  Endorser}: 
An Endorser role is typically implemented by a supply chain entity 
that creates reference Endorsements 
(i.e., claims, values or measurements that are known to be authentic). 
Endorsements contain assertions about the device's intrinsic
trustworthiness and correctness properties. 
Endorsers implement manufacturing, 
productization, or other techniques that establish
the trustworthiness properties of the Attesting Environment.
This is shown as flows~(a) and (b) in
Figure~\ref{fig:TCG-Canonical-Model}.

\item	{\em Owner of Verifier}:  
The Verifier Owner role has policy oversight for the Verifier. 
It generates Appraisal Policy for Evidence and conveys
the policy to the Verifier. 
The Verifier Owner sets policy for acceptable (or unacceptable) Evidence and Endorsements
that may be supplied by Attesters and Endorsers respectively. 

The policies determine the trustworthiness state of the Attester and
how best to represent the state to Relying Parties in the form of Attestation Results.
The Verifier Owner manages Endorsements supplied by Endorsers and may maintain a database of acceptable and/or
unacceptable Endorsements. 
The Verifier Owner authenticates Verifiers and maintains 
lists of trustworthy Endorsers, 
peer Verifiers and Relying Parties with which the Verifier might interact.

\item	{\em Owner of Relying Party}: 

The Relying Party (RP) Owner role has policy oversight for the Relying Party (RP). 
The RP-Owner sets appraisal policy regarding acceptable (or unacceptable) 
Attestation Results about an Attester that was produced by a Verifier. 
The RP-Owner sets appraisal policies on the Relying Party 
that authorizes use of Attestation Results in the context of
the relevant services, management consoles, network equipment, 
an enforcement policies used by the Relying Party. 
The Relying Party Owner authenticates the Relying Party and 
maintains lists of trustworthy Verifiers and peer 
Relying Parties with which the Relying Party might interact.

\item	{\em Evidence}:
The Attestation Evidence is a role message containing assertions from the Attester role. 
Evidence should have freshness and
recentness claims that help establish Evidence relevance. 
For example, a Verifier supplies a nonce that can be
included with the Evidence supplied by the Attester. 
Evidence typically describes the state of the device or entity.
Normally, Evidence is collected in response to a request (e.g. challenge from Verifier). 

Evidence may also describe historical device states
(e.g. the state of the Attester during initial boot). 
It may also describe operational states that are dynamic and
likely to change from one request to the next. 
Attestation protocols may be helpful in providing timing context for
correct evaluation of Evidence that is highly dynamic.

\item	{\em Endorsements}:
Endorsement structures contain reference {\em Claims}
that are signed by an entity performing the Endorser role
(e.g. supply-chain entity or manufacturer of the target device).
Endorsements are reference values that may be used by Owners to form attestation Policies.

Some endorsements may be considered ``intrinsic'' in that
they convey static trustworthiness properties relating to a given actor 
(e.g., device, environment, component, TCB, layer, RoT, or entity).
These may exist as part of the design, implementation, 
validation and manufacture of that actor implementation. 

An Endorser (e.g. manufacturer) may assert immutable and intrinsic claims 
in its Endorsements,
which then allows the Verifier to carry-out appraisal of the Attester (e.g. device)
without requiring Attester reporting
beyond simple authentication.

\end{itemize}

\subsection{Summary of an Attestation Event}
\label{subsec:AttestationEvent}

Figure~\ref{fig:TCG-Canonical-Model} 
illustrates the canonical attestation model. 
When an Attester (e.g.  target device) seeks to 
perform an action at the Relying Party 
(e.g. access resources or services controlled by the Relying Party) 
the Attester must first be evaluated by the Verifier. 
Among its inputs,
the Verifier obtains endorsements 
from the Endorser (e.g. device manufacturer)
in flow~(a) of Figure~\ref{fig:TCG-Canonical-Model}.
Prior to allowing any entity to be evaluated by the Verifier, 
the Owner of the Verifier must first configure 
a number of appraisal policies into the Verifier 
for evaluating Evidences. 
The policies are use-case specific but may require other 
information about the Attester (or User) to be furnished to the Verifier. 
This is shown in Step~1 of Figure~\ref{fig:TCG-Canonical-Model}.
Similarly, in Step~2 the owner of the Relying Party (e.g. resource or service)
must configure a number of Appraisal Policies for Attestation Results into the Relying Party.

When the Attester requests access to the resources at the Relying Party (Step~3),
it will be redirected to the Verifier (Step~4) -- the understanding being
that the Attester must deliver attestation Evidence to the Verifier.
Included here are the endorsement(s) that the Attester obtained previously
from the Endorser (flow(b) of Figure~\ref{fig:TCG-Canonical-Model}). 
The flow represented by Step~3 may be multi-round
and may include a nonce challenge that the Attester
must include in its computation of the Evidence as a means to establish freshness.

After verification and appraisal of the Attester completes, 
the Verifier delivers the Attestation Result to the Relying Party in Step~5.
The Relying Party in its turn must evaluate the Result
against its own policies (set previously in Step~2).
If the Relying Party is satisfied with its evaluation of 
the Attestation Result regarding the Attester,
it will provide the Attester with permission to complete the action
it seeks to perform (e.g. access resources at the RP).

\section{Wallet Attestations to Support VASPs}
\label{sec:AttestWallets}

Following from the previous discussion regarding 
the emerging standard architecture for attestations,
we explore the application of the attestation architecture
for the case of wallet devices.
We consider the potential benefits of wallet attestations
for VASPs, notably in the context of regulated wallets.
Readers interested in the application of attestations to nodes 
(i.e. mining nodes~\cite{Bitcoin}, validator nodes~\cite{Buterin2014})
are directed the work of~\cite{HardjonoSmith2020a}.

In the current work we use the generic term {\em wallet device}
to encompass both the hardware and software
of the wallet system (see the NIST definition of wallets in~\cite{NIST-80202-2018}).
Furthermore,
we use the term broadly to mean wallet systems located
within consumer electronic devices (e.g. mobile devices, smartphones, PC computers, etc.)
as well as Enterprise-grade key management systems
deployed within organizations~\cite{NIST-800-57}.
Thus, as will be seen, attestation capabilities should be used
by wallets regardless of their portability factor (i.e. smartphones)
or legal ownership (i.e. individuals or organizations).

The need to protect keys has been a requirement
since the emergence of digital cryptography.
The need for key protection capabilities expanded with the adoption
of public-key cryptography~\cite{RivestShamirAdleman78} 
into the mainstream computing and networking industry.
In the late 1990s the demand for key protection capabilities
in industry emerged with the rise of the Certification Authorities (CA) business model.
The high-cost of Hardware Security Module (HSM) cards
in the late 1990s meant that only CAs and corporate buyers
could afford these HSM cards.

The effort to produce a low-cost trusted hardware chip commenced
in 1999 with the formation of the Trusted Computing Platform Alliance (TCPA),
which was subsequently renamed the Trusted Computing Group (TCG)~\cite{TCG-website}.
The goal of the TCG was to develop a trusted hardware specifications
that permitted the hardware to be manufactured at very low cost
(e.g. less than a couple of dollars).
The cost had to be extremely low 
-- compared with HSM cards, that could cost several hundred dollars per card --
because the initial targeted market segment was the PC computer market (i.e. PC OEMs),
which is a very cost-sensitive segment of the market.
At the same time, Smart Cards~\cite{RanklEffing2000} were under development
and was targeted primarily for the newly emerging mobile phone market.
Thus the TCG trusted hardware must also be below the cost of smart cards.

The specifications for trusted hardware from the TCG alliance
was called the {\em Trusted Platform Module} (TPM),
with the hardware version {1.2} becoming available in the 2004-2005 timeframe.
Wide deployment of the {TPMv1.2} begun in 2006,
notably with the new purchase requirements from the {U.S.} Army.
More specifically, in February 2006 the {U.S.} Army Small Computer Program published 
a new Consolidated Buy-2 (CB2) Desktop and Notebook minimum specifications for Army customers. 
The Army's new specification requires desktop and laptop personal 
computers be equipped with the new TPM (v1.2) hardware.
This event represented a milestone in the adoption of trusted computing standards.

It is fair to say that the notion of ``attestations'' and device ``measurements''
originated from the TCPA/TCG alliance,
whose members in the early 2000s were grappling 
with the very hard problem of defining {\em technical trust}
(i.e. trust derived from technical means)~\cite{TPM2003Design,Proudler2002}.
Being able to truthfully prove or attest as to the system state
was considered core to the definition and value-proposition of trusted computing.
Unlike HSM cards and Smart Cards,
one of the main driving use-cases for the TPM hardware
was to protect the PC computer,
such as protecting the PC computer through its boot-up sequence (e.g. secure boot).
Thus, the idea is that with the aid of a low-cost well-designed hardware (i.e. the TPM)
that was soldered to the motherboard,
the computer should be able to report (or attest) about its boot-up sequence.
Other applications in the PC context included encrypting files/folders
and self-encrypting disk-drives~\cite{TCG-OPAL-2009,BitlockerTPM}.

\begin{figure}[t]
\centering
\includegraphics[width=0.8\textwidth, trim={0.0cm 0.0cm 0.0cm 0.0cm}, clip]{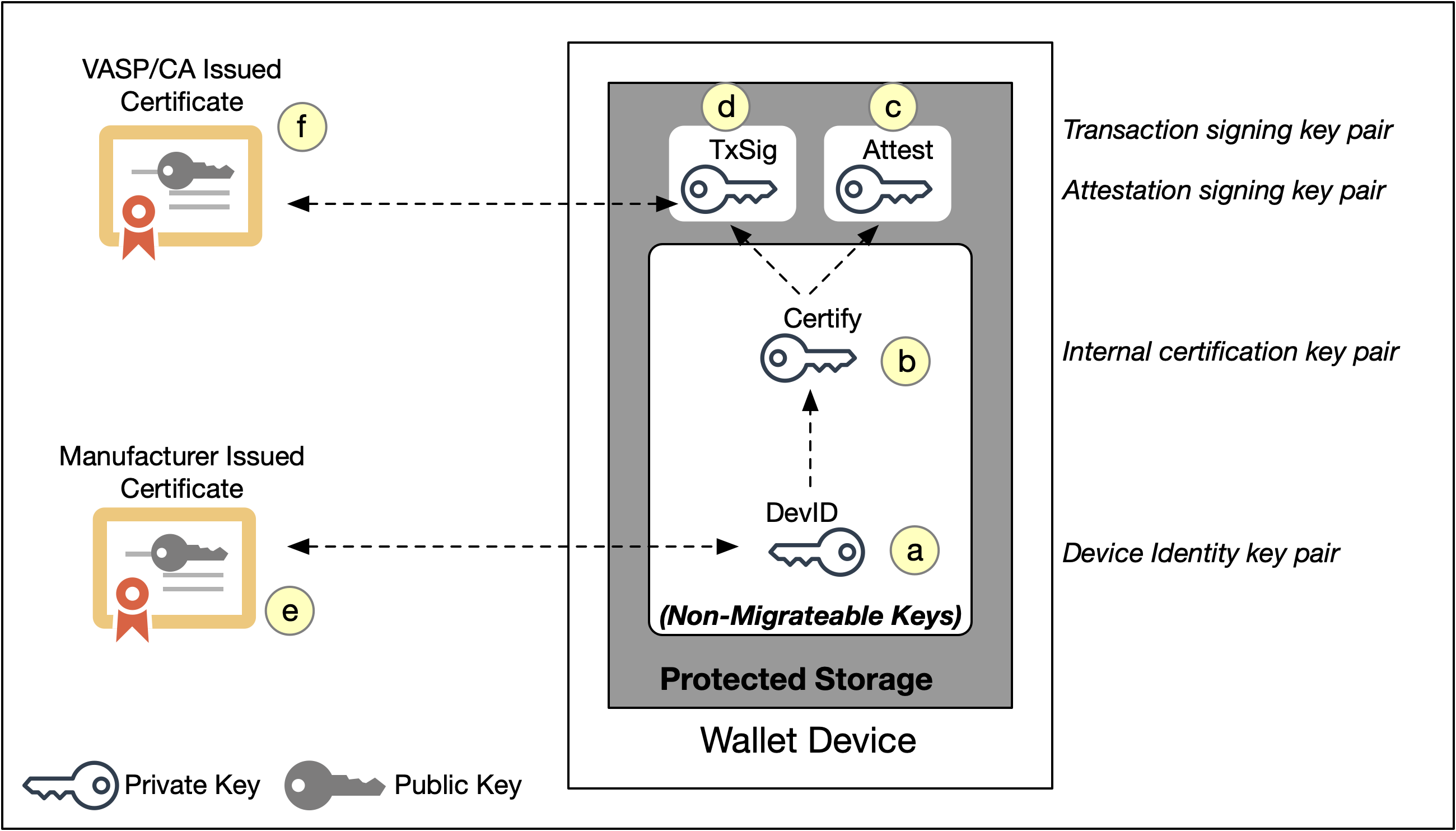}
\caption{High-level illustration of device key hierarchy}
\label{fig:devicekeyhierarchy}
\end{figure}

\subsection{Wallet Devices and Trusted Hardware}

There are several features of trusted hardware
that make it attractive for use in the virtual assets industry~\cite{HardjonoKazmierczak2008b}:
\begin{itemize}

\item	{\em Cryptographic engine, protected storage and tamper-resistance}:
Current trusted hardware typically contains a cryptographic processor which implements
a number of rudimentary functions related to cryptography.
Examples include encryption (symmetric key), digital signatures, hash functions and key-generation.
Trusted hardware typically possess protected storage
for securing keys during system use, and when shutdown.

Tamper-resistance in trusted hardware provides protection against
forced exportation of cryptographic keys (up to a point).
A number of trusted hardware implementation may provide an auto-erasure of keys should
physical tampering occur to sensitive parts in the interior of the hardware.
As such, the value of the asset being protected by the keys should be measured against
the approximate cost of attacking the hardware.

\item	{\em Hardware-bound and non-migratable keys}:
A core feature of trusted hardware such as the TPM
is the ability of certain types of cryptographic keys
to be generated inside the hardware,
and for internal key hierarchies to be established.
Using the example of the TPM,
certain types of keys can be designated as {\em non-migrateable} at creation time,
meaning that the key is bound to that single TPM 
and that it can not be migrated or exported from the TPM (see Figure~\ref{fig:devicekeyhierarchy} (a) and (b)).
The use of non-migratable keys are advantageous 
when addressing the need to prevent the copying of keys.

It is important to note that non-migrateable private-public key pairs
can be used to uniquely identify the device 
(i.e. using the public key)~\cite{TPM1.2specification,IEEE-DeviceID}.
Mechanism to provide privacy to these keys/hardwares
have also been created (see~\cite{Brickell2004,CamenischChen2017}).

\item	{\em Application-level keys linked to non-migratable keys}:
Certain types of keys generated inside the trusted hardware can be designated
to be accessible to application softwares.
Thus,
for the use-case of virtual assets transfers,
one or more key-pairs maybe generated and stored inside the trusted hardware
and be invoked to sign transactions for the blockchain.
The public-key of such key-pairs can be copied to locations external
to the hardware, allowing certificates to be created for that public-key.

A non-migratable key can be used internally to ``certify''
the application-level keys (Figure~\ref{fig:devicekeyhierarchy} (c) and (d)),
thereby providing a provenance link to the non-migratable key (and therefore to trusted hardware).
This feature maybe useful in attestation cases where the user has to prove
the origins of an application-level key-pair.
Typically,
application-level keys can be designated to be {\em migrateable}
at creation time, allowing the key-pair to me migrated (or backed-up)
to a new compatible
trusted hardware using a secure key migration protocol~\cite{TCG-Backup-Migration-2005}.

\item	{\em Hardware-based attestations}:
Certain types of trusted hardware support the truthful reporting of one or more
of its internal state variables,
signed using a reporting-key that is derived from a non-migratable key.
This capability permits an external entity to query
the trusted hardware regarding attributes,
including internal possession of keys (i.e. private-keys),
without revealing the keys.

\end{itemize}
Common examples of trusted hardware that posses some or all of the above features include
the Trusted Platform Module (TPM) hardware 
(version~1.2~\cite{TPM1.2specification} and version~2.0~\cite{TPM2.0specification}),
and the ARM~TrustZone~\cite{ARM-TrustZone2009}.

\begin{figure}[t]
\centering
\includegraphics[width=0.8\textwidth, trim={0.0cm 0.0cm 0.0cm 0.0cm}, clip]{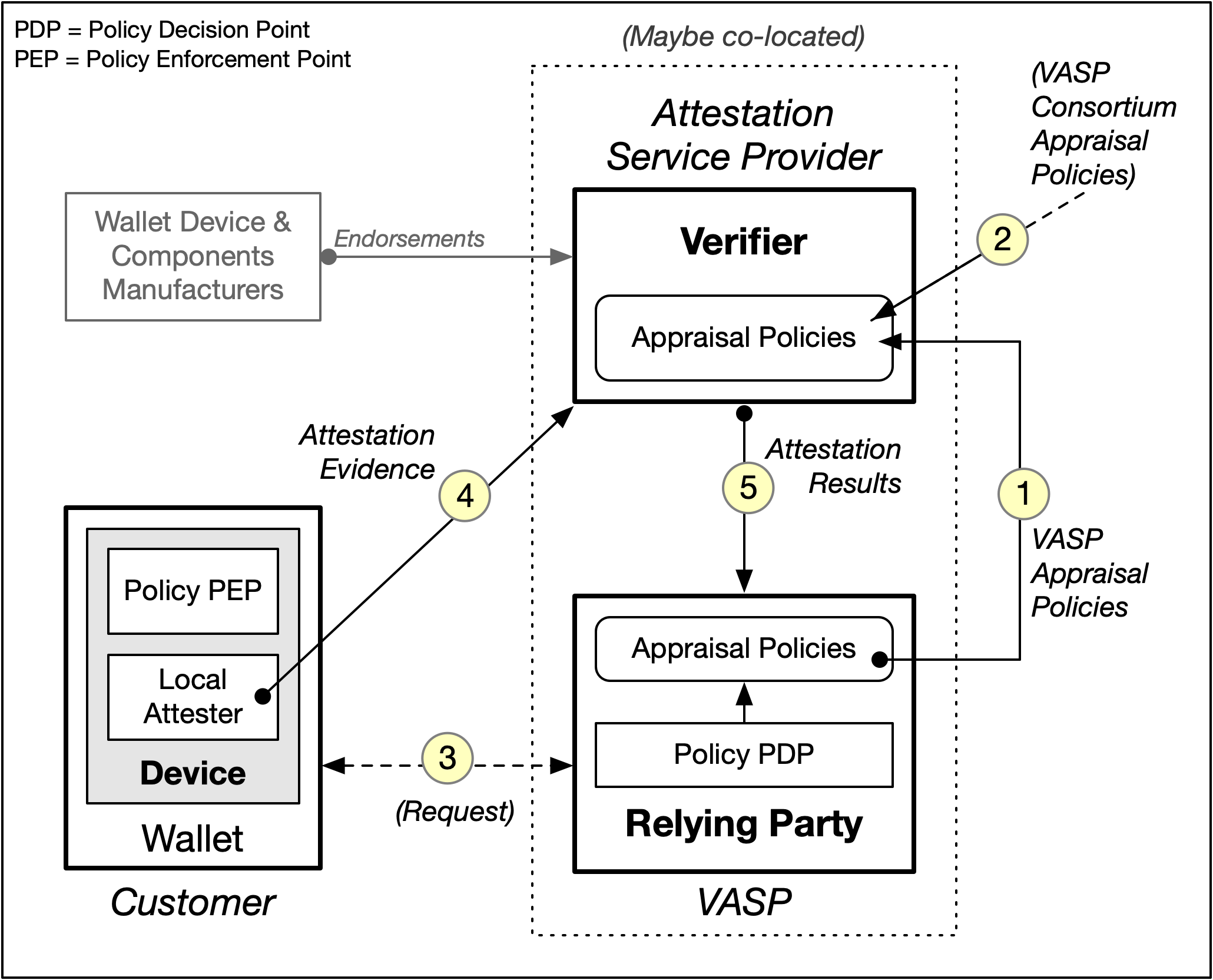}
\caption{Wallet attestation flows}
\label{fig:Walletflow}
\end{figure}

\subsection{Basic Wallet Attestation Flows}

Following from Figure~\ref{fig:TCG-Canonical-Model},
the roles/entities within the wallet attestation flows are as follows 
(see Figure~\ref{fig:Walletflow}):
\begin{itemize}

\item	{\em Appraisal policies defined by VASP}: 
The wallet's attributes of interest to the VASP
can be determined by the VASP configuring the relevant appraisal policies at the ASP (Step~1).
This drives the ASP to request evidences from the wallet device
as appropriate to the configured policies.

In the case that the VASP belongs to a consortium of VASPs (Section~\ref{sec:VASP-networks})
then additional consortium-level policies 
may also be configured at the ASP (Step~2).

\item	{\em Wallet as Attester}: The target device being evaluated 
in this case is the wallet hardware,
as a result of the customer seeking to perform a transaction (Step~3).
The wallet is expected to provide evidence (Step~4), among others, regarding
its installed hardware, software and firmware.

\item	{\em The ASP as Verifier}:
The function of the verifier is represented as the
ASP, which could be a service owned and operated by the VASP,
by a VASP consortium, or by a trusted third party.

\item	{\em The VASP as the Relying Party}:
The relying party in Figure~\ref{fig:Walletflow} is the VASP itself. 
In the regulated-wallet scenario where the VASPs customers
posses wallet devices with trusted hardware,
one goal of the VASP is to obtain attestation-evidence from these devices.
Step~5 illustrates the ASP yielding the attestations results to the VASP as
the relying party.

\end{itemize}

\subsection{Types of Attestation Evidence and Relevance to VASPs}
\label{subsec:VASP-Evidence-Types}

The evidence reportable by trusted hardware
depends largely on the capabilities of the trusted hardware 
and the surrounding system implementing the crypto-wallet functionality.
In general there are a number of system attributes
reportable from a wallet that may complement the customer information
in the context of the Travel Rule.
With regards to key-ownership information and key-operator information
(Section~\ref{subsec:KeyOwnershipInformation}),
attestations technologies allows
a VASP to obtain truthful (unforgeable) information from the wallet,
such as:
(i) {\em how} a key-pair was created (e.g. generated onboard, or injected from outside),
(ii) {\em where} it was created (e.g. under shielded storage),
and 
(iii) {\em the current location} of the key-pair (e.g. geolocation of wallet).

The following is a non-exhaustive list of some of the possible wallet and key information
that can be obtained using attestations:
\begin{itemize}

\item	{\em Key creation provenance}:
Most (if not all) current generation crypto-processor trusted hardware
have the capability to create/generate
a new private-public key pairs inside the protected/shielded location
of the hardware, and to maintain keys inside its long-term non-volatile protected storage.
Furthermore, 
evidence regarding this process can be yielded by the trusted hardware,
allowing the provenance of such keys to be asserted.

Key-provenance evidence is useful for VASPs in many use-case scenarios.
For example,
in the case of a newly on-boarded customer with a wallet, 
the VASP may wish to ascertain the provenance of the existing
customer transaction signing key-pair found in the wallet.
If the provenance of the existing key-pair in the wallet is unverifiable,
then the VASP may require the customer (i.e. wallet)
to generate a new key-pair inside the wallet.

This, in turn,
provides the VASP with a clear line of responsibility and accountability
under the Travel Rule with regards to customer-originated transactions.
The VASP has exculpatory evidence regarding the on-boarding of the new customer
and the start of use of the new key-pair.

\item	{\em Key-type evidence and key loss recovery}:
As mentioned previously,
some crypto-processor trusted hardware (e.g. TPMv1.2 and TPM2.0)
support the creation of non-migratable keys (Figure~\ref{fig:devicekeyhierarchy}).
A VASP may request periodic attestation-evidences from its customers' wallets
regarding the type of the transaction signing key-pair(s) currently in use
(e.g. whether private-key non-migratable).
The VASP may also require these non-migratable keys
to be backed-up to a secure storage location at the VASP, 
using a secure backup/migration protocol
appropriate for the trusted hardware in the wallet~\cite{TCG-Backup-Migration-2005}.
This migration ``blob'' is typically cryptographically sealed in that
it can only be installed onto a new equivalent trusted hardware
under the customer's authorization
(e.g. migration password).
The combined use of key-type evidence and key backup procedure allows VASPs to more effectively
handle emergency cases
(e.g. perceived loss of private-key, actual loss of wallet device, etc.).

For example,
if a wallet device is lost/stolen and the VASP has recent attestation-evidence
that the transaction key is non-migratable,
then this gives the VASP some time to carry out emergency measures
(i.e. assuming it will take some time for the thieves to crack the the tamper-resistance of trusted hardware).
Such emergency procedures, for example,
could mean:
(i) recovering the sealed migration keys (i.e. migration ``blob'') from
the VASP backup storage into a new wallet device with the same/equivalent
trusted hardware;
(ii) activation of trusted hardware and the affected certified transaction signing key-pair;
(iii) moving all asset on the blockchain from that affected public-key (address)
to a new temporary private-public key-pair (e.g. owned by the VASP).

Aside from providing crucial customer service in times of emergency,
the VASP will also obtain sufficient technical evidence 
to justify this customer emergency asset transfer
should the VASP be queried under the Travel Rule
(e.g. why large amounts of assets moved from a customer address to the VASP address).

\item	{\em Evidence of signature-origin of transactions on the blockchain}:
Related to the key creation provenance and key-type,
the use of a hardware-bound private-key to sign transactions
permits the device-origin of that transaction to be ascertained.

This kind of evidence may be important in scenarios
in which the VASP needs proof that a set of confirmed transactions
on the blockchain originated from the specific device
belonging to one of its customers.

\item	{\em Evidence of geolocation of wallet}:
VASPs can obtain evidence regarding the geolocation of a wallet device,
and therefore evidence regarding the geolocation of 
the hardware-bound keys in the wallet.
This may provide a means for VASPs to enforce geolocation-related
policies for customers to ensure that the VASPs customers
are operating within the permissible jurisdiction
(e.g. customer wallet must be in-country to sign transactions).

For example,
the work of~\cite{IETF-draft-ietf-rats-eat-03} includes the ability to
report location coordinates (latitude, longitude and altitude)
of the attester device.
In turn, this can be reinforced with geo-fence policies 
relevant to the specific deployment scenario.

\item	{\em Key usage sequence}:
VASPs can also make use of a number built-in features of trusted hardware
via the application software (e.g. mobile app) on the wallet.
For example,
the application can use the underlying trusted hardware
to maintain a sequential history of the objects (transactions)
signed using the private key inside the trusted hardware
(e.g. in the TPM using the hash-extend operation with
the PCR registers and monotonic counter~\cite{TPM1.2specification,TPM2.0specification}).

For a VASP,
this feature allows the VASP to perform an accurate accounting as to which
order transactions were signed by the wallet system,
as compared to the order in which the transactions were processed 
(i.e. confirmed) on the blockchain,
and whether any transactions were lost
(e.g. transmitted from the wallet, but never reaching the unprocessed-pool 
(UTXO model~\cite{Bitcoin}), etc.).

\item	{\em Evidence of wallet system configurations}:
Attestation technologies allows a VASP
to obtain visibility into the components (hardware, software and firmware)
of the wallets of its customers.

Although this may appear to be intrusive,
this has the advantage of allowing the VASP to advise (or require)
customers to replace a weak wallet system with a stronger system.
More broadly,
this allows VASPs to offer remote device-manageability services to its customers,
including continuous monitoring of the {\em system health} of the wallets.

System health monitoring and reporting
has been deployed in the Enterprise networking industry
for sometime now~\cite{Snyder2006}.
Examples include Microsoft's NAP~\cite{MSFT-NAP-2006},
NAC from Cisco~\cite{Cisco-NAC-Book2007}
and the TNC from the TCG~\cite{TNC2006b,TNC-Arch-2017}.

\end{itemize}

Figure~\ref{fig:devicekeyhierarchy} provides a high level illustration
of a simple key-hierarchy inside the trusted hardware of a wallet system. 
A given device platform may ship with one or more of manufacturer installed keys.
This is shown as the Device Identity Key
in Figure~\ref{fig:devicekeyhierarchy}~(a).
Examples include the manufacturer Endorsement Key (EK) in the TPM~\cite{TCG-TPMkeys-2015},
the DeviceID key in routers~\cite{IEEE-DeviceID,IETF-rats-network-device-attestation-05} and 
the Secret Device Key in server chassis/hardware~\cite{England2016RIOT,Kelly-Cerberus-2017}.
A corresponding certificate
may be issued by the manufacturer (Figure~\ref{fig:devicekeyhierarchy}~(e)).

The customer's transaction signing key-pair (d)
may be derived
from the non-migrateable device identity key (b).
This generational-link between the key (d) to key (b) and to key (a) in 
Figure~\ref{fig:devicekeyhierarchy} allows
the attestation process to discover this link and report it as evidence to the VASP.
Finally,
in Figure~\ref{fig:devicekeyhierarchy}~(f)
the VASP could issue an {X.509} certificate for the customer's
transaction public-key.
The VASP can be the issuing CA, or the VASP can outsource this function
to a commercial CA (see~\cite{HardjonoLipton2020a} for a discussion of VASPs and CAs).
The resulting customer {X.509} certificate could
bear some markings (i.e. specific fields or tags)
indicating that the provenance of the public-key is known to the issuing VASP.
This allows other VASPs and their customers (beneficiaries) to obtain
some degree of confidence regarding the signing key and the wallet system
employed by the originator customer.


\section{Attestation Services within VASP Trust Networks}
\label{sec:VASP-networks}

In order to begin solving the various issues around the Travel Rule and
the challenges in obtaining originator/beneficiary customer information,
the nascent VASP industry should establish
{\em VASP trust networks} or interconnected communities in a manner
similar to the ISP communities on the Internet~\cite{HardjonoLipton2020a}.
The term ``trust network'' is used here to denote
a community of VASPs (e.g. regional level or national level)
that has come together in a consortium arrangement to collaborate under
a common {\em legal trust framework} (system rules definition)
for all its membership.
As an industry consortium, the members of the VASP trust network
would then define the technical specifications and profiles that
would need to be adopted by all members of the consortium.
This ensures a high degree of functional interoperability within
the community of VASPs.

\begin{figure}[t]
\centering
\includegraphics[width=1.0\textwidth, trim={0.0cm 0.0cm 0.0cm 0.0cm}, clip]{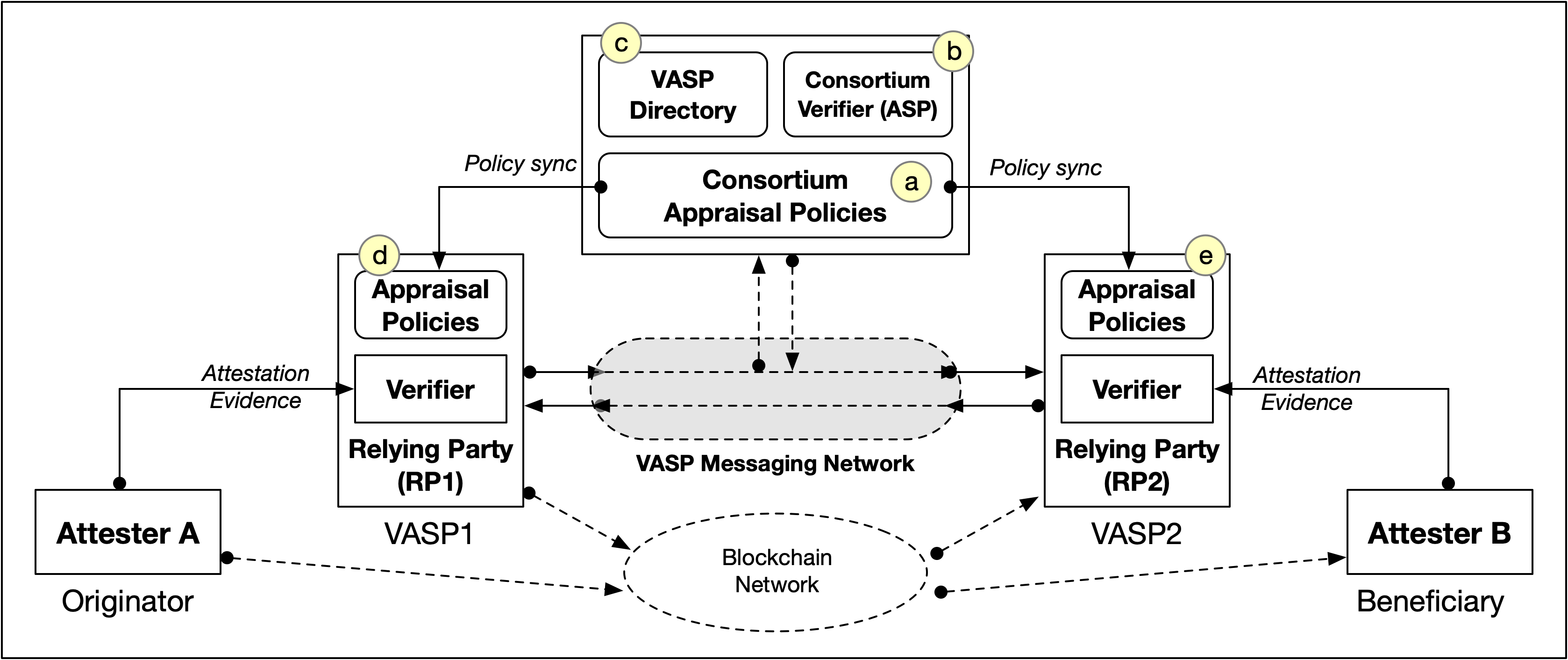}
\caption{Attestation network for VASP consortiums}
\label{fig:VASP-Attestation-Network}
\end{figure}

As part of the service definition of the VASP trust network,
the following services maybe useful to consider in the context
of wallet attestations (Figure~\ref{fig:VASP-Attestation-Network}):
\begin{itemize}

\item	{\em Common baseline appraisal policies}:
The VASP community should develop a set of baseline policies
for the appraisal of wallet systems used in the community.
This may include appraisal policies specific for customer wallets (consumer-grade),
and also appraisal policies for a VASPs key management system,
which could be an enterprise-grade system built also using trusted hardware.
Figure~\ref{fig:VASP-Attestation-Network}(a)
illustrates the notion of a consortium-level appraisal policies.

\item	{\em Common device configuration manifests}:
The community of VASPs should define a number of approved
wallet device configurations (hardware, software, firmware)
in order to allow their customer to obtain
one of the approved configurations.
These approved configuration as defined by their manufacturer
is also known as {\em reference manifests}~\cite{TCG-IWG-2006-Thomas-Ned-Editors-part2,TCG-RIM-2019}.
For each device configuration,
a matching set of appraisal policies should be created
by the consortium and be made accessible to all VASP members.

\item	{\em Shared attestation verification service}:
The consortium as a community should provide
attestation verification services.
This is illustrated in Figure~\ref{fig:VASP-Attestation-Network}(b).
These are also referred to as
Attestation Service Providers (ASP) for the community (see Section~\ref{subsec:TCG-Attestation-entities}).
The work of~\cite{zichardjono2013} points to an example
of an ASP service in the cloud.

\item	{\em Shared integration model for customer single sign-on}:
All VASPs should seek to ensure good customer experiences,
with minimal friction across a variety of wallet application softwares
(e.g. mobile app, dekstop app, browsers, etc).

To this end, the VASP community should harmonize the various
single sign-on (SSO) protocols and flows,
across the various approved types of wallet devices and applications.
Several identity management and SSO protocols 
have been standardized over the past two decades
(e.g. SAML2.0~\cite{SAMLcore,SAMLwebsso}, OAuth2.0~\cite{rfc6749}, OpenID-Connect~\cite{OIDC1.0}).

\item	{\em Cross-VASP attestation verifications}:
A given VASP should posses its own attestation verification system
distinct from the consortium's shared attestation verification service.
This is illustrated in Figure~\ref{fig:VASP-Attestation-Network}(d)-(e).

The goal is to allow a VASP to perform appraisals of evidences
conveyed by the wallet trusted hardware
belonging to a customer of a different VASP (within the consortium).
Having this capability,
the VASP should then be open to appraising evidences from
regulated wallets belonging to customers of non-member VASPs,
and even private-wallets of holders seeking to be on-boarded as new customers.

\end{itemize}

\begin{figure}[t]
\centering
\includegraphics[width=1.0\textwidth, trim={0.0cm 0.0cm 0.0cm 0.0cm}, clip]{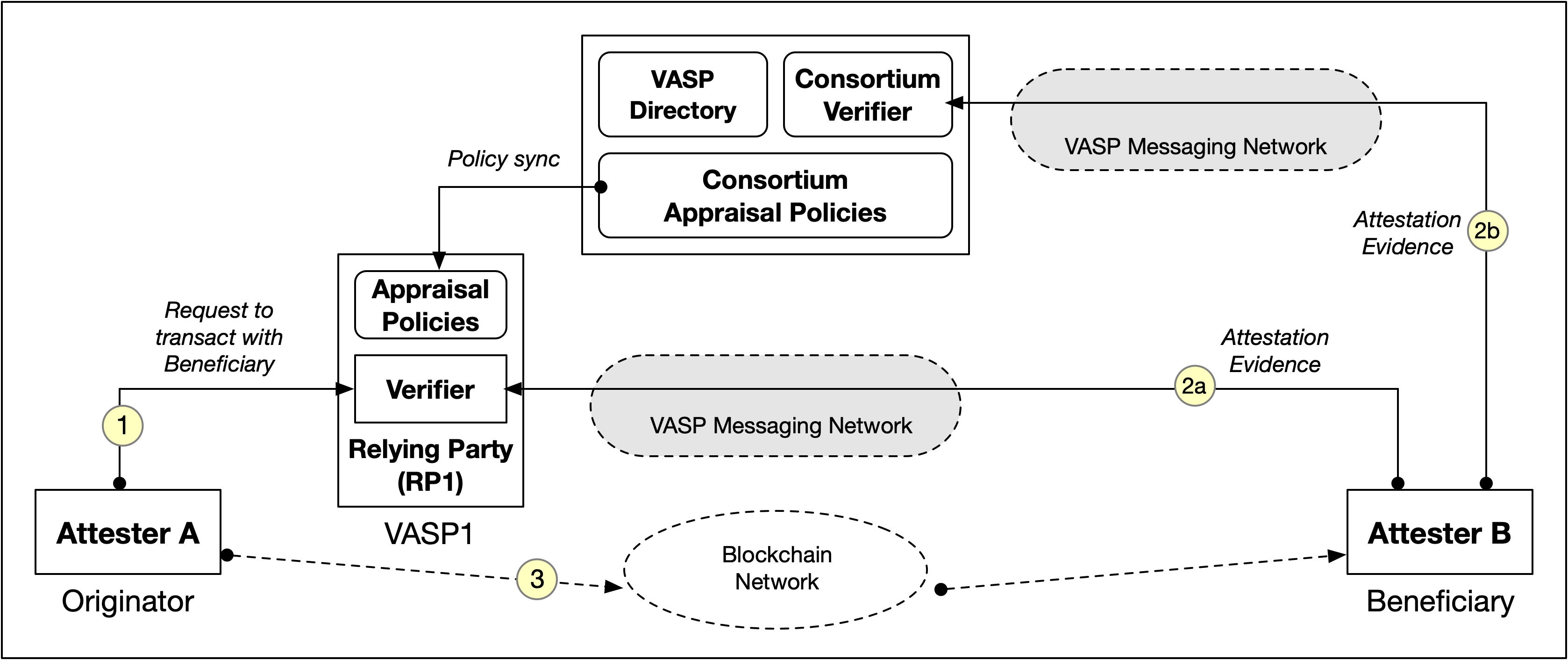}
\caption{Cross-VASP attestation appraisals}
\label{fig:Cross-VASP}
\end{figure}

The notion of Cross-VASP attestation appraisals is summarized in Figure~\ref{fig:Cross-VASP}.
Here,
an originator customer of VASP1 holding
a regulated wallet with trusted hardware seeks
to perform a direct transaction with 
a beneficiary entity who is  customer of VASP2.
Both VASP1 and VASP2 are members of the trust network consortium.
This is shown as Step~(1) of Figure~\ref{fig:Cross-VASP}.
The wallet of the originator is referred to as Attester~A,
while the wallet of the beneficiary is referred to as Attester~B.
Upon request from VASP1,
the beneficiary (Attester~B) generates attestation-evidence
in Step~(2a) and conveys it to the Verifier (ASP) within VASP1.
If both the wallets of the originator and the beneficiary
complies to the appraisal policies of VASP1,
the originator is then given authorization from VASP1
to transact directly with the beneficiary (Step~(3)).
Alternatively,
the VASP1 may require the beneficiary (Attester~B)
to convey its attestation-evidence to the consortium's
ASP, as shown in Step~(2b).

\section{Areas for Innovation}
\label{sec:AreasInnovation}

There are a number of potential areas of innovation
for the newly emerging VASP industry.
Related to the topic of device attestations and trusted hardware
for wallets,
Some of these are as follows
\begin{itemize}

\item	{\em Wallet Levels of Assurance}:
Similar to the notion of {\em Levels of Assurance} (LOA) of 
authentication events defined by NIST~\cite{NIST-800-63-1,OMB-LOA,NIST-800-63b},
a VASP trust network consortium could define
a number of levels assurance as a function of the wallet condition
and other key management aspects.
Today the highest level (Level~4) achievable in the NIST scheme~\cite{NIST-800-63-1}
is one in which a remote network authentication event
employs cryptographic hardware.

For example,
a VASP consortium could recognize a number of different 
types of wallets (e.g. client software, browser plugins, etc.), 
the types of acceptable trusted hardware,
user biometric authentication to the wallet device,
and so on, and use these as input into the wallet LOA matrix.

Using attestations,
wallets could convey evidence to the ASP Service of the VASP consortium
in order to obtain a wallet LOA assignment.
In turn,
this provides some basis for the insurance industry
to begin risk assessment of crypto-asset,
wallets and VASPs.

\item	{\em Wallet LOA for Crypto-Asset Insurance}:
There is interest in the insurance industry to provide insurance services
to crypto-funds~\cite{KharifLouis2018}.
However, the insurance industry will need some technically measurable 
representation of ``trust'' in the VASP management of cryptographic keys.
We believe it is insufficient for VASPs to describe (e.g. in a document) 
the type of trusted hardware 
employed by a VASP and employed by the VASP's customer-wallets.
Instead, attestations evidences should be yielded directly by the wallets
(both enterprise-grade and consumer-grade wallets).

Wallet attestations maybe able to provide insurers
with strong evidence regarding the internal state of the trusted hardware,
the keying material protected by the trusted hardware,
and other aspects of the wallet system.
Crypto-asset insurers may choose to operate its own
evidence verification service (e.g. ASP service)
in order to be able to obtain an independent attestation-result evaluation.

\item	{\em VASP Consortium Repository of Approved Software and Firmware}:
Related to the attestation of wallet devices,
the VASP trust network consortium could maintain a repository
of software, firmware and patches for the various approved wallet devices in it ecosystem.
This could be done in collaboration with the various
software vendors and hardware manufacturers in the wallet space.

Such a repository would assist VASPs in the on-boarding of new customers,
by (i) requiring new customers to obtain 
one of the devices and configurations approved by the VASP trust network consortium,
and (ii) by ensuring that these customer devices
subsequently install only known good software, firmware and patches
from the consortium's repository.

\item	{\em VASP certificate profile for wallet non-migrateable keys}:
The VASP trust network consortium could develop
a certificate profile for transaction signing public-keys
that are provably non-migrateable.

The {X.509} certificate for a transaction public-key (non-migrateable private-key)
could bear some markings that conveys assurance to the recipient
that the corresponding private-key was bound to a trusted hardware.
In turn, this may increase confidence in the counter-party in dealing with the customer
wilding the wallet associated with the {X.509} certificate.

\end{itemize}

\section{Conclusions}
\label{sec:Conclusions}

Today there is a great opportunity for VASPs to shape and influence
the development of future wallets systems with features,
such as attestations,
that would aid VASPs in complying to the the Travel Rule
and help in the manageability of customer wallets.
The ability for a VASP to obtain unforgeable evidence from a wallet system
regarding the provenance of keys,
as well as their usage and location, provides the VASP with additional
means to address the problem of the synchronization between
transaction on the blockchains and the account information required by the Travel Rule.

Wallet attestations maybe able to provide crypto-funds insurers
with strong evidence regarding the key management aspects
of a wallet device,
thereby providing the insurance industry with
measurable levels of assurance that can become the basis
for insurers to perform risk management about a crypto-asset.

Looking more broadly,
if virtual assets and blockchain technology
are going to become a fundamental building block of the new economy
then infrastructures that support the establishment of trust between counter-parties
will need to be developed and deployed.
Because VASPs are the primary touch-points with
users seeking to transact in virtual assets,
this means that deploying trust-infrastructures will be largely the task
of VASPs and the VASP communities globally.

~~\\
~~\\



\end{document}